\documentclass[10pt,aps,prd,amsmath,amssymb,nofootinbib,superscriptaddress,preprintnumbers,hidelinks,
floatfix
]{revtex4-2}

\usepackage{orcidlink}
\usepackage{graphicx} 
\usepackage{dcolumn}
\usepackage{bm}
\usepackage{cancel}
\usepackage{physics}
\usepackage{caption}
\usepackage{booktabs}
\usepackage{multirow}
\usepackage{tensor}
\usepackage{fancyhdr}
\usepackage[normalem]{ulem}
\usepackage{mathtools}

\usepackage{tikz}
\usetikzlibrary{calc}

\newcommand{\ellp}{\ell_\mathrm{P}}
\newcommand{\gnewton}{G_\mathrm{N}}

\newcommand{\linpert}[1]{{#1}_{\mu \nu}^{(1)}}

\numberwithin{equation}{section}
\renewcommand{\theequation}{\arabic{section}.\arabic{equation}}

\begin{document}

\title{\Large \bf  Black hole solutions in\\quantum phenomenological gravitational dynamics}
\author{Ana Alonso-Serrano {\Large \orcidlink{0000-0002-7841-3724}}\,}
\email[]{ana.alonso.serrano@aei.mpg.de}
\affiliation{Institut für Physik, Humboldt-Universität zu Berlin,
Zum Großen Windkanal 6, 12489 Berlin, Germany,
Am Mūhlenberg 1, 14476 Potsdam, Germany}
\affiliation{Max-Planck-Institut für Gravitationsphysik (Albert-Einstein-Institut),
Am Mūhlenberg 1, 14476 Potsdam, Germany}

\author{Marco de Cesare {\Large \orcidlink{0000-0002-4263-1009}}\,}
\email[]{marco.decesare@na.infn.it}
\affiliation{Scuola Superiore Meridionale, Largo S. Marcellino, 10, 80138 Napoli, Italy}
\affiliation{INFN sezione di Napoli, via Cintia, 80126 Napoli, Italy}

\author{Manuel Del Piano {\Large \orcidlink{0000-0003-4515-8787}}\,}
\email[]{manuel.delpiano-ssm@unina.it}
\affiliation{Scuola Superiore Meridionale, Largo S. Marcellino, 10, 80138 Napoli, Italy}
\affiliation{INFN sezione di Napoli, via Cintia, 80126 Napoli, Italy}
\affiliation{Quantum  Theory Center ($\hbar$QTC) \& D-IAS, Southern Denmark University, Campusvej 55, 5230 Odense M, Denmark}

\begin{abstract}
We investigate black hole solutions in a modified gravity theory inspired from a phenomenological approach to quantum gravity based on spacetime thermodynamics developed by Alonso-Serrano and Li{\v s}ka. The field equations are traceless, similarly to unimodular gravity, and include quadratic curvature corrections. We find that static, spherically symmetric, vacuum spacetimes in this theory split into two branches. The first branch is indistinguishable from corresponding solutions in unimodular gravity and describes Schwarzschild-(Anti) de Sitter black holes.
The second branch instead describes horizonless solutions and is characterized by large values of the spatial curvature. We analyze the dynamics of first-order metric perturbations on both branches, showing that there are no deviations from unimodular gravity at this level.
\end{abstract}

\maketitle

\section{Introduction}

A way to approach the search for quantum gravity in recent years has been by analyzing its possible effects as modifications to classical gravitational predictions in experimentally accessible regimes below the Planck energy scale, in what is known as phenomenology of quantum gravity.
As thermodynamics has become an extremely useful tool in gravitational studies, an approach to quantum gravity phenomenology based on spacetime thermodynamics has been proposed in Ref.~\cite{Alonso-Serrano:2020dcz}.
This approach is based on a generalization of Jacobson's derivation of the Einstein field equations from spacetime thermodynamics \cite{Jacobson:1995ab,Jacobson:2003wv,Eling:2006aw}. The generalization is performed by introducing in the thermodynamics input data an extra contribution to the gravitational entropy arising from quantum gravity corrections, which may potentially play a role also below the Planck energy scale. Different approaches to quantum gravity predict logarithmic corrections to the Bekenstein entropy \cite{Kaul:2000kf,Carlip:2000nv,Meissner:2004ju,Sen:2012dw} (as well as to the entanglement entropy), which provide the motivation for this approach.
This results in a
modification to the Einstein field equations including quantum corrections. Interestingly, the resulting field equations are traceless, as in unimodular gravity (for unimodular gravity, see Refs.~\cite{Ellis:2010uc,Carballo-Rubio:2022ofy,Bengochea:2023dep} and references therein).
Previous work analyzed the dynamics of cosmological spacetimes in this theory, showing that there are significant deviations from standard cosmology at early times \cite{Alonso-Serrano:2020dcz,Alonso-Serrano:2022nmb, deCesare:2023fbq, Alonso-Serrano:2023xwr}.

In this work, we perform a full analysis of vacuum, static, spherically symmetric solutions in the theory at hand and examine possible deviations from general relativity. In particular, we show that the Birkhoff theorem does not hold in this theory, due to the branching of the solution space. The dynamics of perturbations around such backgrounds is also analyzed in full detail.

This paper is organized as follows. In Section~\ref{Sec:1} we give a brief review of the field equations of `quantum phenomenological gravitational dynamics'. In Section~\ref{Sec:2} we analyze spherically symmetric static solutions, and show that the solution space splits into two branches. In Section~\ref{Sec:Perturbations}, we derive dynamical equations for perturbations on both branches and compare them with the perturbative dynamics in unimodular gravity and general relativity. A technical appendix is provided, where we review the dynamics of perturbations of Schwarzschild-(Anti) de Sitter geometries in unimodular gravity and general relativity.

\section{Brief Review of Quantum Phenomenological Gravitational Dynamics}\label{Sec:1}

We briefly present the general phenomenological equations of motion, obtained by following similar steps as in the classical derivations of equations of motion in thermodynamics of spacetime. To this end, one first constructs a local observer-dependent horizon in which an equilibrium condition can be imposed.
The total entropy variation, which must be vanishing due to the equilibrium condition, is a sum of the variation of the semiclassical entropy flux of matter fields and the variation of the geometric entropy of the horizon (where the latter is usually interpreted as an entanglement entropy) 
[for a detailed discussion of this construction see Ref.~\cite{Alonso-Serrano:2024amg}]. For our purposes here we directly present the resulting
field equations derived in Ref.~\cite{Alonso-Serrano:2020dcz}
\begin{equation}\label{QPGDeqs}
    \tensor{\mathcal{G}}{_\mu ^\nu}:=\tensor{S}{_\mu ^\nu} - \alpha \kappa S_{\mu \rho}S^{\rho \nu} +\frac{\alpha \kappa}{4}\left( R_{\rho \sigma}R^{\rho \sigma} - \frac{R^2}{4}\right)\tensor{\delta}{_\mu^\nu} = \kappa \left( \tensor{T}{_\mu^\nu} - \frac{T}{4}\tensor{\delta}{_\mu ^\nu} \right)\ , 
\end{equation}
where $S_{\mu \nu} \coloneqq R_{\mu \nu} - R/4 \ g_{\mu \nu}$ is the traceless part of the Ricci tensor, $T_{\mu \nu}$ is the Hilbert stress-energy tensor defined through the matter Lagrangian (for further details on the explicit derivation see, e.g.~Refs.~\cite{Alonso-Serrano:2020dcz,Alonso-Serrano:2024amg,Jacobson:2018ahi,Iyer:1996ky}).
The quadratic curvature terms in the field equations, parametrized by the dimensionless constant $\alpha$,\footnote{Here we use $\kappa = 8 \pi \gnewton$, with $c=1$, such that $\alpha \kappa = D \ellp^2$ from Eq.~(3.43) in Ref.~\cite{Alonso-Serrano:2020dcz} and $\alpha$ includes a $\hbar$ factor. become relevant as the Planck scale is approached. These terms stem directly from logarithmic corrections in the entropy, as shown in Ref.~\cite{Alonso-Serrano:2020dcz}. 
The actual numerical value of $\alpha$ depends on the particular approach to quantum gravity considered. Therefore, for the sake of generality, $\alpha$ will be treated as a free parameter in this work.}
Note that Eq.~\eqref{QPGDeqs} is traceless and for $\alpha=0$ reduces to the field equations of unimodular gravity, which is recovered when no quantum gravity corrections are introduced.

Taking the covariant divergence of both sides of Eq.~\eqref{QPGDeqs} and using the contracted Bianchi identities $\nabla^\nu (R_{\mu\nu}-\frac{1}{2}R\, g_{\mu\nu})=0$~, we obtain \cite{Alonso-Serrano:2020dcz}
\begin{equation}\label{Eq:Bianchi}
    \frac{1}{4}\nabla_\mu R-\alpha \kappa \nabla_\nu\left( S_{\mu \rho}S^{\rho \nu} \right)+\frac{\alpha \kappa}{4}\nabla_\mu\left( R_{\rho \sigma}R^{\rho \sigma} - \frac{R^2}{4}\right)=\kappa\left(\nabla_\nu \tensor{T}{_\mu^\nu}-\frac{1}{4}\nabla_\mu T \right)~.
\end{equation}
Unlike general relativity, but similarly to unimodular gravity (whose field equations are also traceless), conservation of the stress-energy tensor of matter fields does not follow from the Bianchi identities, but constitutes an independent assumption \cite{Alonso-Serrano:2020dcz,Ellis:2010uc}. We will work under this assumption, i.e.~$\nabla^\nu T_{\mu\nu}=0$, throughout the paper.\footnote{In the context of quantum gravity phenomenology, the possibility that spacetime discreteness may lead to the non-conservation of the matter stress-energy tensor has been proposed in Refs.~\cite{Josset:2016vrq,Perez:2017krv}. This scenario can be consistently embedded into unimodular gravity, and some of its consequences for cosmology have been explored in Refs.~\cite{Perez:2020cwa,deCesare:2021wmk,Landau:2022mhm,Zhai:2025hfi}. Generalizations of this proposal to the model at hand with field equations \eqref{Eq:Bianchi}, though possible in principle, go beyond the scope of this work, where we focus on vacuum solutions.}
We note that Eq.~\eqref{Eq:Bianchi} has a more complicated structure than its counterpart in unimodular gravity, particularly due to the second term on the left-hand side, which for generic spacetime geometries cannot be expressed as the gradient of a scalar function. Nonetheless, for some symmetric spacetimes, Eq.~\eqref{Eq:Bianchi}  can still be integrated explicitly. In these cases, the cosmological constant arises as an integration constant, in complete analogy with unimodular gravity \cite{Ellis:2010uc}. This property has been discussed for cosmological spacetimes in Refs.~\cite{Alonso-Serrano:2020dcz,deCesare:2023fbq}. In the next section we show that a similar feature also holds for some vacuum spherically symmetric spacetimes.

\section{Static Vacuum Solutions in Spherical Symmetry}\label{Sec:2}

Let us consider a general ansatz for a static and spherically symmetric metric 
\begin{equation}\label{Eq:MetricAnsatz}
    \dd s^2 = - e^{2 \nu(r)} F(r)\,  \dd t^2 + \frac{\dd r^2}{F(r)} + r^2\, \dd \Omega^2 \ ,
\end{equation}
where $r$ is the areal radius and $\dd \Omega^2 = \dd \theta^2 + \sin^2 \theta \, \dd \varphi^2$ denotes the round metric on the unit 2-sphere.
    
Substituting \eqref{Eq:MetricAnsatz} into the field equations \eqref{QPGDeqs} and assuming a vacuum spacetime (i.e., $T_{\mu\nu}=0$) we obtain the following set of dynamical equations
\begin{subequations}
\begin{align}
    4 r^3 \tensor{\mathcal{G}}{_t^t} & =  2 \alpha \kappa \, F \, \nu^{\prime}  \left(  r F \nu^{\prime} -  \mathcal{A} \right) - r\,\mathcal{A}  - 4 r^2 F  \nu^{\prime}  = 0 \ , \label{Eq: eq1} \\
    4 r^3 \tensor{\mathcal{G}}{_r^r} & =  2 \alpha \kappa \, F \, \nu^{\prime}  \left( r F \nu^{\prime} + \mathcal{A} \right)  - r\,\mathcal{A}  + 4 r^2 F  \nu^{\prime}  = 0 \ , \label{Eq: eq2} \\
    4r^2 \tensor{\mathcal{G}}{_\theta ^\theta} &= \mathcal{A}+2 \alpha  \kappa  F^2 \nu^{\prime \, 2} = 0  \ , \label{Eq: eq3} \\
\tensor{\mathcal{G}}{_\varphi ^\varphi} &= \tensor{\mathcal{G}}{_\theta^\theta} = 0  \ , \label{Eq: eq4}
\end{align}
\end{subequations}
where, for notational convenience, we introduced the auxiliary function
\begin{equation}
    \mathcal{A}  \coloneqq   r^2 F^{\prime \prime} + 3 r^2 F^{\prime } \nu^{\prime} + 2F \left( r^2 \nu^{\prime \prime} +  r^2 \nu^{\prime \, 2}-1\right) + 2 ~.
\end{equation}
Throughout the paper, a prime denotes derivative with respect to the radial coordinate $r$. To make the notation lighter, the functional dependence of $F$ and $\nu$ on $r$ is implied. Due to the symmetries of the line element \eqref{Eq:MetricAnsatz}, the remaining components of $\tensor{\mathcal{G}}{_\mu^\nu}$ are identically vanishing. It is evident that the equations above are not all independent.
In fact, taking linear combinations  of Eqs.~\eqref{Eq: eq1} and \eqref{Eq: eq2}, we obtain
\begin{subequations}
    \begin{align}
      -2r^2 \left( \tensor{\mathcal{G}}{_r^r} +\tensor{\mathcal{G}}{_t^t}\right)&=  {\cal A}  +2\alpha\kappa F^2\nu^{\prime\,2} = 0 ~,\label{eq:second_combination}\\
       r^3 \left( \tensor{\mathcal{G}}{_r^r} - \tensor{\mathcal{G}}{_t^t}\right)&=  \left( 2 r^2 + \alpha  \kappa  \, \mathcal{A} \right) F \, \nu^{\prime}= 0 ~.
       \label{eq: branching equation}
    \end{align}
\end{subequations}
We observe that Eq.~\eqref{eq:second_combination} is identical to the angular equation \eqref{Eq: eq3}. Hence, Eqs.~\eqref{eq:second_combination}, \eqref{eq: branching equation} represent a minimal set of independent equations for the system at hand.

Given the factorized form of equation~\eqref{eq: branching equation},
we can identify two distinct branches, depending on whether $\nu^{\prime}$ or the combinations of terms in brackets vanish. Therefore, the Birkhoff theorem does not hold in this theory. The two branches are analyzed separately in the following subsections. As shown below, solutions on the first branch do not entail any departures from general relativity, while solutions on the second branch are characterized by significant $\alpha$-dependent corrections.

\subsection{Branch I}\label{Sec:BranchI}
On this branch, Eq.~\eqref{eq: branching equation} is satisfied by requiring $\nu^\prime=0$~.
This implies $\nu(r) = \nu_0$~, with $\nu_0$ a constant.
Substituting into Eq.~\eqref{Eq: eq1}, we obtain
\begin{equation}\label{Eq: branch 1 eq F}
    r^2 F^{\prime \prime} - 2 F + 2 = 0 ~,
\end{equation}
whose general solution is
\begin{equation}\label{Eq: branch 1 F sol}
F(r)=1-\frac{2GM}{r}-\frac{\Lambda}{3}r^2,
\end{equation}
where the mass $M$ and the cosmological constant $\Lambda$ arise as free integration constants. The fact that $\Lambda$ arises as an integration constant is a basic consequence of the field equations being trace-free---in contrast with GR, where it is introduced at the level of the gravitational action. This is analogous to unimodular gravity \cite{Ellis:2010uc}, and a similar property has been already observed for cosmological solutions in the theory at hand \cite{Alonso-Serrano:2020dcz,deCesare:2023fbq}.

Thus, solutions on this branch are Schwarzschild-(Anti) de Sitter black holes. We also observe that such solutions do not display any dependence on $\alpha$ and are therefore indistinguishable from corresponding solutions in general relativity. For this reason, in the following we will also refer to this branch as the `GR branch' (since the dynamics is identical to GR, even though the cosmological constant has a different nature here, as discussed above).  

\subsection{Branch II}\label{Sec:BranchII}

Next, we consider the case in which $\nu^\prime \neq 0$.
Then, in order to satisfy Eq.~\eqref{eq: branching equation}
we demand that the combination of terms enclosed in brackets vanish, that is
\begin{equation}\label{Eq: branch 2}
    2 r^2 + \alpha  \kappa  \, \mathcal{A}= 0 ~.
\end{equation}
A simple equation for $\nu$ can be obtained combining Eqs.~\eqref{Eq: branch 2} and Eq.~$\eqref{Eq: eq3}$ to eliminate ${\cal A}$
\begin{equation}\label{Eq:nu_branch2}
   (\alpha\kappa)^2 F^2 \nu^{\prime \,2} - r^2 = 0 ~.
\end{equation}
This is solved by quadrature
\begin{equation}\label{Eq: branch 2 nu prime}
        \nu(r) = \frac{\sigma}{\alpha \kappa} \int\dd r\,\frac{r}{F(r)} \qq{where} \sigma \coloneqq \pm 1 \ .
\end{equation}
Substituting Eq.~\eqref{Eq: branch 2 nu prime} into Eq.~\eqref{Eq: eq2} we obtain a second order non-linear ordinary differential equation (ODE) for $F(r)$
\begin{equation}\label{Eq: branch 2 F}
    (\alpha \kappa )^2 r^2 F \, F^{\prime \prime}+ \sigma \, \alpha  \kappa  \, r^3  F^{\prime }+ 2 F \left[ (\alpha  \kappa )^2+ (1+\sigma )\alpha  \kappa  \, r^2 \right]-2 (\alpha \kappa )^2 F^2+2 r^4 = 0\ .
\end{equation}

An exact solution to Eq.~\eqref{Eq: branch 2 F} is
\begin{equation}\label{Eq: branch 2 F dS}
    F(r) = 1 + \frac{r^2}{\alpha \kappa} ~, \quad \mbox{with $\sigma=-1$}~.
\end{equation}
Substituting the solution \eqref{Eq: branch 2 F dS} into Eq.~\eqref{Eq: branch 2 nu prime}, we obtain
\begin{equation}\label{Eq: branch 2 nu dS}
    \nu(r) = -\frac{1}{2}\log\left( 1 +\frac{r^2}{\alpha \kappa} \right) + \nu_0 \ ,
\end{equation}
where $\nu_0$ is an integration constant. Inserting the solutions \eqref{Eq: branch 2 F dS} and \eqref{Eq: branch 2 nu dS} into the line element \eqref{Eq:MetricAnsatz}, we obtain 
\begin{equation}\label{Eq: branch 2 new line el}
    \dd s^2 = - e^{2\nu_0} \dd t^2 + \left( 1 +\frac{r^2}{\alpha \kappa} \right)^{-1} \dd r^2 + r^2 \dd \Omega^2 ~.
\end{equation}
Rescaling the time coordinate as $ t \mapsto e^{-\nu_0} \, t$ and introducing a new coordinate $\chi$~, defined as
$\chi=\arcsin{(r/r_0)}$ for $\alpha<0$ and $\chi= {\rm arcsinh}(r/r_0)$ for $\alpha>0$~, with $r_0\coloneqq\sqrt{|\alpha|\kappa}$~, we can recast \eqref{Eq: branch 2 new line el} as
\begin{subequations}\label{Eq:PlanckianBackgrounds}
\begin{align}
\dd s^2 &= -  \dd t^2 + r_0^2\left(\dd \chi^2 +  \sin^2\chi \,\dd \Omega^2 \right) ~,\quad \mbox{for $\alpha<0$}~,\label{Eq:PlanckianBackgrounds_1}\\
\dd s^2 &= -  \dd t^2 + r_0^2\left( \dd \chi^2 +  \sinh^2\chi\, \dd \Omega^2 \right) ~,\quad \mbox{for $\alpha>0$}~.\label{Eq:PlanckianBackgrounds_2}
\end{align}
\end{subequations}
These are static spacetimes whose constant-$t$ slices have the topology of a 3-sphere and a 3-hyperboloid, respectively, both with spatial curvature radius $r_0$~.
We note that, if $\alpha \sim \order{1}$~, $r_0$ is of the order of the Planck length $\ell_{\rm Pl}=\sqrt{\kappa}$~. It is interesting to observe that the solutions \eqref{Eq:PlanckianBackgrounds_1}, \eqref{Eq:PlanckianBackgrounds_2}   correspond to `frozen' Friedmann-Lemaître-Robertson-Walker (FLRW) universes with a constant scale factor, with closed and open topologies, respectively. Their stability will be investigated in Section~\ref{Sec:Perturbations_BranchII}.

Interestingly, the solution \eqref{Eq: branch 2 F dS} is not the only solution of Eq.~\eqref{Eq: branch 2 F} that admits a power-law expansion in $r$
\begin{equation}\label{Eq:PowerSeries_F_BranchII}
    F(r) = \sum_{n=-L}^\infty  a_n \left(\frac{r}{\sqrt{|\alpha|\kappa}}\right)^{n} ~,
\end{equation}
with $L$ a finite and positive integer. Substituting \eqref{Eq:PowerSeries_F_BranchII} into Eq.~\eqref{Eq: branch 2 F}, we obtain a tower of
algebraic non-linear equations for the expansion coefficients $a_n$. We find that the coefficients $a_{-k}=0=a_{2k+1}$
(with $k$ any positive integer)
are all vanishing for either cases $\sigma=\pm 1$. In particular, since $a_{-1}$ is vanishing, this prevents us from obtaining Schwarzschild-like solutions with this ansatz. On the one hand, in the case of $\sigma=1$, we obtain a unique solution, for which we provide the first few expansion coefficients: $a_0 = 0 = a_2, \  a_4=-1, \ a_6 = 4 , \ a_8 = -25 , \ a_{10} =226~$. On the other hand, fixing $\sigma=-1$, we obtain two solutions. The first one is given by the coefficients: $a_0 = 0 = a_2, \  a_4=-1, \ a_6 = -2 , \ a_8 = -11 , \ a_{10} = -82  \ \ldots$~, while the second and less trivial solution is given by
\begin{gather*}
    a_0 = 1~, \quad a_4=\frac{a_2-1}{5}~, \quad  a_6 = \frac{7a_2-5a_2^2-2}{70} ~, \quad a_8=\frac{70a_2^3 -127a_2^2+77 a_2-20}{1890} ~,\\ \ a_{10} = \frac{-9450a_2^4 + 21110a_2^3 - 19091 a_2^2 + 8857a_2-1426}{415800} ~ .
\end{gather*}
We remark that the latter solution is parametrized by the coefficient $a_2$, which cannot be determined by Eq.~\eqref{Eq: branch 2 F}. Making a particular choice $a_2=1$, all of the higher-order coefficients vanish, and the solution \eqref{Eq: branch 2 F dS} is recovered.

More in general, it is easy to show that Branch II solutions do not describe black holes, in that such solutions do not have an event horizon. In fact, assuming that $F(r)$ has a zero at $r=\tilde{r}\neq0$, Eq.~\eqref{Eq: branch 2 F} implies $F^{\prime}(\tilde{r})=-2\sigma \tilde{r}/(\alpha \kappa)$. Substituting this result into Eq.~\eqref{Eq: branch 2 nu prime} we obtain, in the proximity of $\tilde{r}$, $\nu(r)\approx-\frac{1}{2}\log(r/\tilde{r}-1)+C$, where $C$ is an integration constant. Therefore, we have $g_{tt}(\tilde{r})= 2\sigma \tilde{r}^2\exp(2C)/(\alpha\kappa)\neq0$, which implies that there is no horizon at $r=\tilde{r}$. For this kind of solutions, Eq.~\eqref{Eq: branch 2 F} can be solved systematically by means of a power series expansion around $\tilde{r}$. Such solutions are analytic around $\tilde{r}$ and read as
\begin{equation}\label{Eq:solutionF_analytic}
    F(r)= -\frac{2\sigma\tilde{r}}{\alpha\kappa}(r-\tilde{r})-\left(\frac{2}{\tilde{r}^2}+\frac{2+\sigma}{\alpha\kappa} \right)(r-\tilde{r})^2+\mathcal{O}(r-\tilde{r})^3~.
\end{equation}
Thus, for either sign of $\sigma$, \eqref{Eq:solutionF_analytic} gives a one-parameter family of solutions parametrized by $\tilde{r}$.
The above results also show that such solutions do not have a well-defined $\alpha\to0$ limit.

\section{Dynamics of perturbations}\label{Sec:Perturbations}

In this Section we analyze the dynamics of first-order perturbations on both branches. The perturbed metric is
\begin{equation}\label{Eq: metric perturb}
    \overline{g}_{\mu \nu} \coloneqq g_{\mu \nu} + h_{\mu \nu}~,
\end{equation}
where $g_{\mu \nu}$ is the background metric and $h_{\mu\nu}$ denotes metric perturbations. Since we are interested in vacuum backgrounds, the first nontrivial contributions of the stress-energy tensor of matter fields arise at the perturbative level. We assume that metric and matter perturbations are of the same order, that is $|h_{\mu \nu}|\,,~|T_{\mu \nu}| \sim \order{\varepsilon}$ with $\varepsilon \ll 1$~.
In the following, we make a notational shift: we replace all quantities in Eq.~\eqref{QPGDeqs} with their barred counterparts and proceed expanding perturbatively in $\varepsilon$. From now on, unbarred quantities will refer to the background. Furthermore, indices will be raised using the background metric. Following this scheme, the field equations \eqref{QPGDeqs} can be expanded perturbatively to first order in $\varepsilon$. In the following subsections, we will study the dynamics of linear perturbations around background solutions on both branches.

\subsection{Branch I perturbations}\label{Sec:Perturbations_BranchI}
As shown in Section~\ref{Sec:BranchI}, Branch I background solutions are Schwarzschild-(Anti) de Sitter black holes.
Expanding the field equations \eqref{QPGDeqs} around such a background, we obtain to first order in $\varepsilon$
\begin{equation}\label{Eq: perturb}
    S^{(1)}_{\mu \nu} = \kappa \left( T_{\mu \nu} - \frac{T}{4}g_{\mu \nu}\right) \ , 
\end{equation}
where
\begin{equation}\label{Eq: perturbed s}
   S^{(1)}_{\mu \nu} = \Lambda \left( h_{\mu \nu} - \frac{h}{4} g_{\mu \nu}\right) + \frac{1}{2}\left( \nabla_\mu \nabla_\nu \, h + \Box h_{\mu \nu} - \nabla_\mu \nabla^\lambda \, h_{\nu \lambda} - \nabla_\nu \nabla^\lambda \, h_{\mu \lambda}\right) + \frac{1}{4}g_{\mu \nu} \left( \nabla^\lambda \nabla^\rho h_{\lambda \rho} - \Box h\right)  \ ,
\end{equation}
is the first-order perturbation of $S_{\mu \nu}$ and where $\Box \coloneqq g^{\mu \nu} \nabla_\mu \nabla_\nu$~, $h\coloneqq g^{\mu \nu} h_{\mu \nu} $. 
Note that the perturbed field equations \eqref{Eq: perturbed s} are trace-free with respect to the background $g_{\mu \nu}$~: this property is inherited from the full field equations \eqref{QPGDeqs}.

Let us remark that the perturbative equations \eqref{Eq: perturbed s} do not involve any $\alpha$-dependent terms. This property has already been observed in Section~\ref{Sec:BranchI} for background solutions on Branch I, and the equation above shows that it also holds for perturbations around such backgrounds.
Moreover, since $\alpha$-corrections enter the non-linear field equations \eqref{QPGDeqs} only through terms which are quadratic in the Ricci curvature, the equations for linear perturbations coincide with those of unimodular gravity. (For ease of comparison, the dynamical equations for perturbations in general relativity and unimodular gravity are reviewed in the Appendix.)
Therefore, we conclude that the dynamics of first-order perturbations
on this branch is insensitive to quantum-gravity motivated corrections in the present model---irrespective of the mass of the black hole and the value of the cosmological constant.

\subsection{Branch II perturbations}\label{Sec:Perturbations_BranchII}
Moving on to Branch II backgrounds, we take the exact solution~\eqref{Eq: branch 2 new line el} as a background (after rescaling the time coordinate as $ t \mapsto e^{-\nu_0} \, t$ for convenience). Notice that, unlike Branch I solutions, such a background is not an Einstein manifold, that is $R_{\mu \nu}\neq k\,  g_{\mu \nu}$ and therefore, the simplifications leading  from Eq.~\eqref{QPGDeqs} to Eq.~\eqref{Eq: perturb} do not take place in this case.
Instead, we have $\tensor{R}{_0^0} = 0$ and $\tensor{R}{_i^j} = -\frac{2}{\alpha\kappa} \tensor{\delta}{_i^j}$, where $i,j$ denote spatial indices.

In order to investigate how $\alpha$-corrections enter the linearized dynamics on this Branch, let us examine axial and polar metric perturbations separately. For simplicity, in this section we focus on vacuum perturbations (i.e.~we assume $T_{\mu\nu}=0$).

\subsubsection{Axial perturbations}
In the Regge-Wheeler gauge \cite{Regge:1957td,Wald:1984rg,Maggiore:2018sht,Chandrasekhar:1985kt}
and assuming a monochromatic axial perturbation with frequency $\omega$, the perturbed line element reads
\begin{equation}
    \dd s^2 = - \dd t^2 + \frac{\dd r^2}{1+\frac{r^2}{\alpha\kappa}} + r^2 (\dd \theta^2 + \sin^2\theta\, \dd \varphi^2) + 2 e^{-i \omega t}\sin \theta \,\dv{P_\ell(\cos \theta)}{\theta} \Big( h_0(r) \dd t + h_1(r) \dd r\Big) \dd \varphi \ ,
\end{equation}
where $|h_0(r)|, |h_1(r)| \sim \order{\varepsilon}$ are the radial profiles of the linear axial perturbations, and $P_\ell(\cos \theta)$ is the Legendre polynomial of degree $\ell$. Expanding the field equations~\eqref{QPGDeqs} to linear order, we find that the angular and radial parts factorize (as in general relativity), and the radial profile of the perturbations obeys the following system of first order ODEs
\begin{align}
    &i \omega  \, h_0^\prime- 2 i \omega  \, \frac{h_0}{r} + \left(\frac{\ell(\ell+1)-2}{r^2}-\omega^2\right)h_1 =0 ~,\label{Eq: pert branch 2 eq1}\\
    &i \omega \, h_0 + \frac{r}{\alpha\kappa}  \, h_1 +  \left(1 +\frac{ r^2}{\alpha\kappa}\right) h_1^\prime =0 ~. \label{Eq: pert branch 2 eq2}
\end{align}
This result shows that also in this case there are no contributions coming from $\alpha$-dependent corrections.
Equations~\eqref{Eq: pert branch 2 eq2} and \eqref{Eq: pert branch 2 eq1} can be combined to yield a single second-order ODE for $h_1$
\begin{equation}\label{Eq: pert h1 eq}
   \left(1 +\frac{r^2}{\alpha \kappa}\right)h_1^{\prime \prime} - \left(2+\frac{r^2}{\alpha \kappa}\right)\frac{h_1^{\prime}}{r} - \left[\frac{\ell(\ell+1)-2}{r^2} - \left(\omega^2-\frac{1}{\alpha\kappa}\right)\right]h_1 = 0 \ .
\end{equation}

Now, we notice that Eq.~\eqref{Eq: pert h1 eq} may be cast in a Schrödinger-like form by introducing the following auxiliary function $Q(r)$ as 
\begin{equation}
    h_1(r) = \frac{r\, Q(r)}{\sqrt{1+\frac{r^2}{\alpha\kappa}}} \ , 
\end{equation}
and transforming the radial coordinate as
\begin{equation}\label{rastdef}
    \dd r_\ast = \frac{\dd r}{\sqrt{1+\frac{r^2}{\alpha\kappa}}} \ .
\end{equation}
With this transformation, Eq.~\eqref{Eq: pert h1 eq} reads
\begin{equation}\label{Eq:SchrodingerForm}
     \dv[2]{Q}{r_\ast} + \left[ \omega^2 - \left(\frac{\ell(\ell+1)}{r^2} + \frac{1}{\alpha\kappa} \right)  \right]Q = 0\ .
\end{equation}
In the equation above, $r$ should be regarded as a function of $r_\ast$. Note that at large distances the effect of the centrifugal barrier is negligible; therefore, the solution is either purely oscillating or exponentially damped/undamped depending on the sign of $\omega^2-1/(\alpha\kappa)$.
Once Eq.~\eqref{Eq:SchrodingerForm} has been solved for $Q(r)$, Eq.~\eqref{Eq: pert branch 2 eq2} can be solved to determine $h_0$ as
\begin{equation}
    h_0(r)=\frac{i}{\omega}\sqrt{1+\frac{r^2}{\alpha\kappa}} \,\big(r \, Q(r) \big)^\prime~.
\end{equation}

\subsubsection{Polar perturbations}
We can extend the analysis above to monochromatic polar perturbations of the metric. In the Regge-Wheeler gauge~\cite{Regge:1957td},
the perturbed metric reads
\begin{align}
    \dd s^2 = -\big(1+ H_0(r)A(t,\theta)\big)\dd t^2 + \frac{1 + H_2(r)A(t,\theta) }{1+\frac{r^2}{\alpha\kappa}}\dd r^2 +  2 H_1(r) A(t,\theta)  \, \dd t \, \dd r + r^2\big(1+ K(r) A(t,\theta) \big)\big(\dd \theta^2 + \sin^2 \theta \, \dd \varphi^2\big) \ ,
\end{align}
where we defined $A(t,\theta)\coloneqq e^{-i \omega t}P_\ell(\cos \theta) $ as a shorthand notation and $|H_0(r)|, |H_1(r)|, |H_2(r)|, |K(r)| \sim\order{\varepsilon}$. Substituting this metric into the field equations \eqref{QPGDeqs} and expanding to first order in $\varepsilon$, we find the dynamical equations for polar perturbations. Specifically, we obtain the algebraic constraint equation $H_2(r)=H_0(r)$ and the following set of equations for the remaining dynamical variables
\begin{subequations}
\begin{align}
 &i \omega  H_1 - (H_0 + K)^\prime = 0 \ ,\label{Eq:Polar1} \\
 &\left(1+\frac{r^2}{\alpha\kappa}\right) \left( H_0^{\prime\prime}-\frac{2 K^{\prime}}{r} - 2 i \omega H_1^{\prime}\right)+ \frac{r}{\alpha\kappa} H_0^\prime - \left( \omega^2+\frac{2}{\alpha\kappa} + \frac{2}{r^2}\right) H_0 - \frac{2i  \omega}{\alpha\kappa}   r H_1+ \frac{\ell(\ell+1)-2}{r^2} K =0 \ ,\label{Eq:Polar2} \\
 &\left(1+\frac{r^2}{\alpha\kappa} \right)\left(\frac{2 i \omega H_1}{r} -K^{\prime \prime}\right) + \left(\frac{2 K^{\prime}}{r}+ \frac{3r}{\alpha \kappa}\right) K^{\prime} -  \omega^2 K  + \left(\frac{\ell (\ell+1)}{r^2}+\frac{2}{\alpha\kappa} \right) H_0 =0 \ .\label{Eq:Polar3} 
\end{align}
\end{subequations}
We note that the parameter $\alpha$ only enters the dynamical equations for perturbations through the background curvature $\sim 1/(\alpha\kappa)$.
The above can be combined to yield
\begin{equation}\label{Eq:PolarSchrodinger}
    \dv[2]{W}{r_\ast} + \left[ \omega^2 - \left(\frac{\ell(\ell+1)}{r^2} + \frac{1}{\alpha\kappa}\right)  \right]W = 0\ , 
\end{equation}
where we introduced the quantity $W(r)\coloneqq (H_0(r)+K(r))/r$ and $r_\ast$ is defined as in \eqref{rastdef}. When written in this form, the equation for polar perturbations \eqref{Eq:PolarSchrodinger} exactly matches that for axial perturbations for the variable $Q$, Eq.~\eqref{Eq:SchrodingerForm}. Once Eq.~\eqref{Eq:PolarSchrodinger} has been solved, Eq.~\eqref{Eq:Polar1} can be solved by quadrature to determine $H_1$. 
Lastly, another linearly independent combination of the system above gives the following dynamical equation for $H_0$, where $W$ plays the role of a source
\begin{equation}
\begin{split}
    \left(1 + \frac{r^2}{\alpha\kappa}\right)H_0^{\prime\prime} + \frac{2+ \frac{3r^2}{\alpha\kappa}   }{r}H_0^{\prime} - \left( \frac{\ell (\ell+1)}{r^2} + \omega ^2+\frac{2r^2}{\alpha\kappa} \right)H_0 =
   \left(1 + \frac{r^2}{\alpha\kappa}\right) \frac{6 W'(r)}{r}
   +\left(\frac{\ell(\ell+1)-2}{r^2}- 2 \omega ^2\right)W(r)~.
   \end{split}
\end{equation}

Lastly, we observe that, since equations \eqref{Eq:SchrodingerForm} and \eqref{Eq:PolarSchrodinger} have identical form, the axial and polar sectors for perturbations trivially have the same quasi-normal mode spectrum \cite{ChandrasekahrQNMs}. Similar properties are well-known to hold for black hole backgrounds in general relativity \cite{Chandrasekhar:1985kt,Pound:2021qin,Sasaki:2003xr}.

\section{Conclusion}
We showed that, in spherical symmetry, static vacuum solutions of quantum gravitational phenomenological dynamics split into two branches. Branch I solutions are Schwarzschild-(Anti) de Sitter spacetimes. As such, they are indistinguishable from corresponding black hole solutions in general relativity. On the other hand, Branch II solutions do not describe black holes. Rather, they are horizonless spacetimes and are characterized by large values of the curvature. We obtained an exact solution on this branch, which describes a non-expanding FLRW spacetime with non-zero spatial curvature, of the order of the Planck curvature scale $\sim \ell_{Pl}^{-2}$. A full classification of Branch II solutions is left for future work.

We showed in full generality that perturbations of Branch I backgrounds do not entail any deviations from the perturbative dynamics of unimodular gravity, regardless of the black hole mass and the value of the cosmological constant. For Branch II, we focused on the exact background solution discussed above and considered vacuum gravitational perturbations. We analyzed separately the dynamics of both polar and axial metric perturbations. Also in this case, there are no deviations from unimodular gravity for either polarity sector of the perturbations.

Our results show that, under the above assumptions, the solutions of the field equations \eqref{QPGDeqs} either coincide with spherically symmetric black hole solutions in GR (Branch I), or describe exotic geometries without an event horizon and asymptotic flatness (Branch II).
One of the key expectations from a consistent theory of quantum gravity is the resolution of spacetime singularities. 
For instance, within the context of loop quantum gravity, the gravitational interaction becomes repulsive at the Planck scale. This causes collapsing matter to rebounce when the energy density approaches a critical value~\cite{Husain:2021ojz,Han:2023wxg}. Quantum geometry effects also lead to singularity resolution in vacuum~\cite{Ashtekar:2018lag,Bodendorfer:2019cyv,Alonso-Bardaji:2021yls}.
On the other hand, in an effective approach, singularity resolution may also be achieved by introducing a matter stress-energy tensor with exotic properties such that gravity becomes repulsive at small length scales. This approach has been pursued by several authors \cite{Hayward:2005gi,Dymnikova:1992ux,Simpson:2018tsi,Simpson:2019mud,Bardeen:1968zz,AyonBeato:1998ub}. 
In contrast, as shown in this work, no such mechanisms exist for singularity resolution for static and spherically symmetric black hole geometries in the present model in the absence of matter fields.
Future work should investigate whether this result stems from the idealized restriction to the vacuum and spherically symmetric case, and study more physically realistic configurations beyond spherical symmetry and in the presence of matter fields.

\section*{Acknowledgements}
AA-S is funded by the Deutsche Forschungsgemeinschaft (DFG, German Research Foundation) — Project ID 51673086, and acknowledges support through Grants No.~PID2020–118159 GB-C44 and PID2023-149018NB-C44 (funded by MCIN/AEI/10.13039/501100011033). MdC acknowledges support from INFN iniziativa specifica GeoSymQFT. MDP acknowledges support from INFN iniziativa specifica ST\&FI and expresses gratitude to the Southern Denmark University, the Danish Institute for Advanced Study, and the Quantum Theory Center in Odense (Denmark) for hosting during the final stages of the work. This work contributes to COST Action CA23130 -- Bridging high and low energies in search of quantum gravity (BridgeQG). 

\appendix

\section*{Appendix: Perturbations of Schwarzschild-(Anti) de Sitter in general relativity and unimodular gravity}\label{Sec:Perturbations_GR_UG}

\renewcommand{\theequation}{A.\arabic{equation}}

In this Appendix we review the dynamics of linear metric and matter perturbations in general relativity and unimodular gravity, around the same Schwarzschild-(Anti) de Sitter background.

The perturbative field equations in general relativity read as
\begin{equation}\label{Eq:GR_perturbation}
2\linpert{G} = \nabla_\mu \nabla^\rho h_{\nu \rho} + \nabla_\nu \nabla^\rho h_{\mu \rho} - \Box h_{\mu \nu} - \nabla_\mu \nabla_\nu h + \left(\Lambda \, h + \Box h - \nabla^\rho \nabla^\sigma h_{\rho \sigma}\right) g_{\mu \nu} - 2 \Lambda \, h_{\mu \nu} 
    = 2 \kappa \, T_{\mu \nu} \ , 
\end{equation}
where $\Box \coloneqq g^{\mu \nu} \nabla_\mu \nabla_\nu$, $h\coloneqq g^{\mu \nu} h_{\mu \nu} $.
Upon contraction with the inverse background metric $g^{\mu \nu}$~, Eq.~\eqref{Eq:GR_perturbation} yields
\begin{equation}
    G^{(1)}= \Box h -\nabla^\rho \nabla^\sigma h_{\rho \sigma} + \Lambda \, h = \kappa \, T \ .
\end{equation}
The divergence of the l.h.s.~of Eq.~\eqref{Eq:GR_perturbation} is identically zero, $\nabla^{\mu}\linpert{G}=0$~, which is a linearized version of the Bianchi identities. Therefore, taking the divergence of both sides of Eq.~\eqref{Eq:GR_perturbation} we obtain $\nabla^{\mu}T_{\mu \nu}=0$~.

On the other hand, the field equations of unimodular gravity read \cite{Perez:2020cwa}
\begin{equation}
    S_{\mu \nu} \coloneqq R_{\mu \nu} - \frac{R}{4} g_{\mu \nu} = \kappa \left( T_{\mu \nu} - \frac{T}{4}g_{\mu \nu} \right) ~.
\end{equation}
Upon inserting the perturbed metric \eqref{Eq: metric perturb} and linearizing the field equations around the Schwarzschild-(Anti) de Sitter background, we obtain the perturbative equations
\begin{equation}\label{EQ:perturb_UG}
2\linpert{S} = \nabla_\mu \nabla^\rho h_{\nu \rho} + \nabla_\nu \nabla^\rho h_{\mu \rho} - \Box h_{\mu \nu}- \nabla_{\mu} \nabla_\nu h   + \frac{1}{2}\left( \Box h - \nabla^\rho \nabla^\sigma h_{\rho \sigma} + \Lambda \, h  \right) g_{\mu \nu} -  2 \Lambda \, h_{\mu \nu}= 2\kappa\left(T_{\mu \nu}-\frac{T}{4} g_{\mu\nu}\right) \ .
\end{equation}
Contracting these equations with the inverse background metric, it can be easily realized that they are traceless. In fact, taking the divergence of both sides in Eq.~\eqref{EQ:perturb_UG}, we obtain
\begin{equation}\label{EQ:div_perturb UG}
\nabla^\nu\left(\linpert{S} + \frac{\kappa}{4}T ~ g_{\mu \nu}  \right) = \kappa ~ \nabla^\nu T_{\mu \nu} ~.
\end{equation}
Using the linearized Bianchi identity and $\linpert{S} = \linpert{G} - \frac{1}{4}G^{(1)}g_{\mu \nu}$~, this equation boils down to
\begin{equation}
 -\frac{1}{4}\nabla_\mu\left(G^{(1)} -  \kappa \, T \right) =  \kappa ~ \nabla^\nu T_{\mu \nu} ~ . 
\end{equation}
If we further assume the  conservation of the stress-energy tensor of matter fields (i.e., $\nabla^\nu T_{\mu \nu} =0$, which constitutes an independent assumption in unimodular gravity), we finally obtain
\begin{equation}\label{Eq:IntegrationConstUG}
    G^{(1)} -  \kappa \, T = c~,
\end{equation}
where $c$ is an integration constant. Finally, using Eq.~\eqref{Eq:IntegrationConstUG} it is straightforward to show that Eq.~\eqref{EQ:perturb_UG} can be recast in the form \eqref{Eq:GR_perturbation} with a shifted cosmological constant $\Lambda\mapsto\Lambda+c$~.

\bibliographystyle{unsrt}
\bibliography{biblio.bib}

\begin{thebibliography}{10}

\bibitem{Alonso-Serrano:2020dcz}
Ana Alonso-Serrano and Marek Li\v{s}ka.
\newblock {Quantum phenomenological gravitational dynamics: A general view from
  thermodynamics of spacetime}.
\newblock {\em JHEP}, 12:196, 2020.

\bibitem{Jacobson:1995ab}
Ted Jacobson.
\newblock {Thermodynamics of space-time: The Einstein equation of state}.
\newblock {\em Phys. Rev. Lett.}, 75:1260--1263, 1995.

\bibitem{Jacobson:2003wv}
Ted Jacobson and Renaud Parentani.
\newblock {Horizon entropy}.
\newblock {\em Found. Phys.}, 33:323--348, 2003.

\bibitem{Eling:2006aw}
Christopher Eling, Raf Guedens, and Ted Jacobson.
\newblock {Non-equilibrium thermodynamics of spacetime}.
\newblock {\em Phys. Rev. Lett.}, 96:121301, 2006.

\bibitem{Kaul:2000kf}
Romesh~K. Kaul and Parthasarathi Majumdar.
\newblock {Logarithmic correction to the Bekenstein-Hawking entropy}.
\newblock {\em Phys. Rev. Lett.}, 84:5255--5257, 2000.

\bibitem{Carlip:2000nv}
Steven Carlip.
\newblock {Logarithmic corrections to black hole entropy from the Cardy
  formula}.
\newblock {\em Class. Quant. Grav.}, 17:4175--4186, 2000.

\bibitem{Meissner:2004ju}
Krzysztof~A. Meissner.
\newblock {Black hole entropy in loop quantum gravity}.
\newblock {\em Class. Quant. Grav.}, 21:5245--5252, 2004.

\bibitem{Sen:2012dw}
Ashoke Sen.
\newblock {Logarithmic Corrections to Schwarzschild and Other Non-extremal
  Black Hole Entropy in Different Dimensions}.
\newblock {\em JHEP}, 04:156, 2013.

\bibitem{Ellis:2010uc}
George F.~R. Ellis, Henk van Elst, Jeff Murugan, and Jean-Philippe Uzan.
\newblock {On the Trace-Free Einstein Equations as a Viable Alternative to
  General Relativity}.
\newblock {\em Class. Quant. Grav.}, 28:225007, 2011.

\bibitem{Carballo-Rubio:2022ofy}
Ra\'ul Carballo-Rubio, Luis~J. Garay, and Gerardo Garc\'\i{}a-Moreno.
\newblock {Unimodular gravity vs general relativity: a status report}.
\newblock {\em Class. Quant. Grav.}, 39(24):243001, 2022.

\bibitem{Bengochea:2023dep}
Gabriel~R. Bengochea, Gabriel Leon, Alejandro Perez, and Daniel Sudarsky.
\newblock {A clarification on prevailing misconceptions in unimodular gravity}.
\newblock {\em JCAP}, 11:011, 2023.

\bibitem{Alonso-Serrano:2022nmb}
Ana Alonso-Serrano, Marek Li\v{s}ka, and Antonio Vicente-Becerril.
\newblock {Friedmann equations and cosmic bounce in a modified cosmological
  scenario}.
\newblock {\em Phys. Lett. B}, 839:137827, 2023.

\bibitem{deCesare:2023fbq}
Marco de~Cesare and Giulia Gubitosi.
\newblock {Cosmological evolution from modified Bekenstein entropy law}.
\newblock {\em JCAP}, 01:046, 2024.

\bibitem{Alonso-Serrano:2023xwr}
Ana Alonso-Serrano, Guillermo A.~Mena Marug\'{a}n, and Antonio
  Vicente-Becerril.
\newblock Analytic primordial power spectrum in the dressed metric approach to
  loop quantum cosmology and thermodynamics of spacetime.
\newblock {\em International Journal of Modern Physics D}, 34(01):2450062,
  2025.

\bibitem{Alonso-Serrano:2024amg}
Ana Alonso-Serrano, Luis~J. Garay, and Marek Li\v{s}ka.
\newblock {From spacetime thermodynamics to Weyl transverse gravity}.
\newblock {\em Phys. Rev. D}, 111(6):064019, 2025.

\bibitem{Jacobson:2018ahi}
Ted Jacobson and Manus Visser.
\newblock {Gravitational Thermodynamics of Causal Diamonds in (A)dS}.
\newblock {\em SciPost Phys.}, 7(6):079, 2019.

\bibitem{Iyer:1996ky}
Vivek Iyer.
\newblock {Lagrangian perfect fluids and black hole mechanics}.
\newblock {\em Phys. Rev. D}, 55:3411--3426, 1997.

\bibitem{Josset:2016vrq}
Thibaut Josset, Alejandro Perez, and Daniel Sudarsky.
\newblock {Dark Energy from Violation of Energy Conservation}.
\newblock {\em Phys. Rev. Lett.}, 118(2):021102, 2017.

\bibitem{Perez:2017krv}
Alejandro Perez and Daniel Sudarsky.
\newblock {Dark energy from quantum gravity discreteness}.
\newblock {\em Phys. Rev. Lett.}, 122(22):221302, 2019.

\bibitem{Perez:2020cwa}
Alejandro Perez, Daniel Sudarsky, and Edward Wilson-Ewing.
\newblock {Resolving the $H_0$ tension with diffusion}.
\newblock {\em Gen. Rel. Grav.}, 53(1):7, 2021.

\bibitem{deCesare:2021wmk}
Marco de~Cesare and Edward Wilson-Ewing.
\newblock {Interacting dark sector from the trace-free Einstein equations:
  Cosmological perturbations with no instability}.
\newblock {\em Phys. Rev. D}, 106(2):023527, 2022.

\bibitem{Landau:2022mhm}
Susana~J. Landau, Micol Benetti, Alejandro Perez, and Daniel Sudarsky.
\newblock {Cosmological constraints on unimodular gravity models with
  diffusion}.
\newblock {\em Phys. Rev. D}, 108(4):043524, 2023.

\bibitem{Zhai:2025hfi}
Yuejia Zhai, Marco de~Cesare, Carsten van~de Bruck, Eleonora Di~Valentino, and
  Edward Wilson-Ewing.
\newblock {A low-redshift preference for an interacting dark energy model}.
\newblock 3 2025.

\bibitem{Regge:1957td}
Tullio Regge and John~A. Wheeler.
\newblock {Stability of a Schwarzschild singularity}.
\newblock {\em Phys. Rev.}, 108:1063--1069, 1957.

\bibitem{Wald:1984rg}
Robert~M. Wald.
\newblock {\em {General Relativity}}.
\newblock Chicago Univ. Pr., Chicago, USA, 1984.

\bibitem{Maggiore:2018sht}
Michele Maggiore.
\newblock {\em {Gravitational Waves. Vol. 2: Astrophysics and Cosmology}}.
\newblock Oxford University Press, 3 2018.

\bibitem{Chandrasekhar:1985kt}
Subrahmanyan Chandrasekhar.
\newblock {\em {The mathematical theory of black holes}}.
\newblock {(Clarendon Press, Oxford, 1985)}, 1985.

\bibitem{ChandrasekahrQNMs}
S.~Chandrasekhar and S.~Detweiler.
\newblock The quasi-normal modes of the schwarzschild black hole.
\newblock {\em Proceedings of the Royal Society of London. Series A,
  Mathematical and Physical Sciences}, 344(1639):441--452, 1975.

\bibitem{Pound:2021qin}
Adam Pound and Barry Wardell.
\newblock {\em Black Hole Perturbation Theory and Gravitational Self-Force},
  pages 1411--1529.
\newblock Springer Nature Singapore, Singapore, 2022.

\bibitem{Sasaki:2003xr}
Misao Sasaki and Hideyuki Tagoshi.
\newblock {Analytic black hole perturbation approach to gravitational
  radiation}.
\newblock {\em Living Rev. Rel.}, 6:6, 2003.

\bibitem{Husain:2021ojz}
Viqar Husain, Jarod~George Kelly, Robert Santacruz, and Edward Wilson-Ewing.
\newblock {Quantum Gravity of Dust Collapse: Shock Waves from Black Holes}.
\newblock {\em Phys. Rev. Lett.}, 128(12):121301, 2022.

\bibitem{Han:2023wxg}
Muxin Han, Carlo Rovelli, and Farshid Soltani.
\newblock {Geometry of the black-to-white hole transition within a single
  asymptotic region}.
\newblock {\em Phys. Rev. D}, 107(6):064011, 2023.

\bibitem{Ashtekar:2018lag}
Abhay Ashtekar, Javier Olmedo, and Parampreet Singh.
\newblock {Quantum Transfiguration of Kruskal Black Holes}.
\newblock {\em Phys. Rev. Lett.}, 121(24):241301, 2018.

\bibitem{Bodendorfer:2019cyv}
Norbert Bodendorfer, Fabio~M. Mele, and Johannes M{\"u}nch.
\newblock {Effective Quantum Extended Spacetime of Polymer Schwarzschild Black
  Hole}.
\newblock {\em Class. Quant. Grav.}, 36(19):195015, 2019.

\bibitem{Alonso-Bardaji:2021yls}
Asier Alonso-Bardaji, David Brizuela, and Ra{\"u}l Vera.
\newblock {An effective model for the quantum Schwarzschild black hole}.
\newblock {\em Phys. Lett. B}, 829:137075, 2022.

\bibitem{Hayward:2005gi}
Sean~A. Hayward.
\newblock {Formation and evaporation of regular black holes}.
\newblock {\em Phys. Rev. Lett.}, 96:031103, 2006.

\bibitem{Dymnikova:1992ux}
I.~Dymnikova.
\newblock {Vacuum nonsingular black hole}.
\newblock {\em Gen. Rel. Grav.}, 24:235--242, 1992.

\bibitem{Simpson:2018tsi}
Alex Simpson and Matt Visser.
\newblock {Black-bounce to traversable wormhole}.
\newblock {\em JCAP}, 02:042, 2019.

\bibitem{Simpson:2019mud}
Alex Simpson and Matt Visser.
\newblock {Regular black holes with asymptotically Minkowski cores}.
\newblock {\em Universe}, 6(1):8, 2019.

\bibitem{Bardeen:1968zz}
James~M. Bardeen.
\newblock Non-singular general-relativistic gravitational collapse.
\newblock In {\em Proceedings of the International Conference GR5}, page 174,
  1968.
\newblock Tbilisi, USSR.

\bibitem{AyonBeato:1998ub}
Eloy Ayón-Beato and Alberto García.
\newblock Regular black hole in general relativity coupled to nonlinear
  electrodynamics.
\newblock {\em Phys. Rev. Lett.}, 80:5056--5059, 1998.

\end{thebibliography}
\end{document}